\documentclass[12pt, letterpaper]{article}
\usepackage[utf8]{inputenc}
\usepackage[utf8]{inputenc}
\usepackage[english]{babel}
\usepackage{amsmath}
\usepackage{amsfonts}
\usepackage{setspace}
\usepackage[ruled,vlined,linesnumbered,resetcount]{algorithm2e}
\usepackage[margin=1in]{geometry}
\usepackage{natbib}
\usepackage{caption}
\usepackage{graphicx}
\usepackage{setspace}
\setcitestyle{authoryear}
\usepackage{authblk}

\title{The Importance of Social and Government Learning in Ex Ante Policy Evaluation}

\author[1]{Gonzalo Casta\~neda}
\author[2,3]{Omar A. Guerrero}
\affil[1]{Centro de Investigación y Docencia Econ\'omica (CIDE), Mexico City}
\affil[2]{The Alan Turing Institute, London}
\affil[3]{Department of Economics, University College London, London}
\date{}

\begin{document}

\maketitle

\begin{abstract}
We provide two methodological insights on \emph{ex ante} policy evaluation for macro models of economic development. First, we show that the problems of parameter instability and lack of behavioral constancy can be overcome by considering learning dynamics. Hence, instead of defining social constructs as fixed exogenous parameters, we represent them through stable functional relationships such as social norms. Second, we demonstrate how agent computing can be used for this purpose. By deploying a model of policy prioritization with endogenous government behavior, we estimate the performance of different policy regimes. We find that, while strictly adhering to policy recommendations increases efficiency, the nature of such recipes has a bigger effect. In other words, while it is true that lack of discipline is detrimental to prescription outcomes (a common defense of failed recommendations), it is more important that such prescriptions consider the systemic and adaptive nature of the policymaking process (something neglected by traditional technocratic advice). 
\end{abstract}

\section{Introduction}\label{sec:introduction}

The process of prioritizing public policies for economic development involves a high-dimensional topical space with a vast set of potential priorities to be explored. This leads to the question of how to proceed with \emph{ex ante} policy evaluations through a rigorous macro model. In the economic literature, there are two major concerns on whether this can be done. The first one is the popular `Lucas critique' \citep{lucas_econometric_1976} and the second (and earlier) critique comes from \cite{mises_ultimate_1962} and several members of the Austrian school.

According to the Lucas critique, parameters estimated from a reduced-form model are directly related to the agents' optimal decision rules and, indirectly, to the policies prevailing during the sampling period. Consequently, when counterfactual policies are analyzed with these estimates, there is no guarantee of parameter invariance since the individuals' expectations and their corresponding decisions change.  On the other hand, some members of the Austrian school, argue that theories are not falsifiable through empirical evidence because human actions lack constancy. Since humans' purposes and motivations are not observed, and no physiological explanation can be offered, nothing can be said about how a person might react to certain policy. According to \cite{israel_modern_2016}, human actions are contingent to their knowledge and beliefs. Hence, their learning process prevents the constancy principle required for the use of statistical relationships in the prescription and evaluation of public policies.

In our view, contemporary economists are much better equipped to deal with the Lucas critique and Mises' constancy problem. On one hand, advances in experimental and behavioral economics, psychology and neuroeconomics provide a better understanding of human motivations and learning mechanisms \citep{bowles_moral_2017,dhami_foundations_2016}. On the other, computational methods allow us to explicitly model adaptive processes such as the one through which governments reorganize their priorities as a response to policy outcomes.

In this paper, we differentiate social learning from two other popular channels of parameter instability: \emph{expectations} and \emph{social preferences}. Social learning is particularly salient in development economics, especially when the problem at hand involves corruption and budgetary allocations across different policy issues. We say that a \emph{social learning channel} exists when the implementation of the policy induces certain collective behavior that, in turn, `nudges' individuals to conform with a norm. The latter impinges on the policymaking process through the government's response (adaptation) to policy outcomes. 

Clearly, \emph{ex ante} evaluation is subject to how well economic models can account for collective behavior. For this reason, we advocate for the use of agent computing. In order to demonstrate the adequacy of this tool, we infer policy priorities and evaluate different prescriptions through a computational model by \cite{castaneda_how_2018}. The model posits a political economy game where a government allocates resources to different policy issues, updating the allocation profile while a norm of corruption emerges among the officials in charge of implementing the public policies.

Our empirical results show that, on average, policies derived from this computational framework are better, in terms of gains in efficiency (\emph{i.e.,} lower corruption), than those neglecting social learning and endogenous government behavior. While this, of course, is not the only factor that improves efficiency, it is seems to be more important than the commonly used argument of being disciplined when adopting policy prescriptions. The rest of the paper is organized as follows. In section \ref{sec:modelling}, we discuss the problem of parameter instability through a social learning channel and the importance of modeling endogenous responses. Section \ref{sec:model} presents the model's equations. Section \ref{sec:method} introduces the methods and metrics to be used in the empirical analysis. In section \ref{sec:empirics}, we show and interpret our main empirical findings. Finally, section \ref{sec:discussion} presents some reflections on policy formulation.

\section{On econometric models and policy evaluation}\label{sec:modelling}

Criticisms of econometric policy evaluation with observational data have traditionally focused on the problem of parameter instability. This means that policy advice derived from statistical relationships is not valid when the socioeconomic environment has changed. The sources of such instability are multiple. First, it may result from drastic exogenous shocks (\emph{e.g.}, hikes in energy and food prices, or natural disasters) that affect the environment where agents make decisions. Second, it can be produced by endogenous perturbations to the system such as a financial crises, technological breakthroughs, institutional adjustments, cultural changes or any type of structural transformation. Third, it could be a consequence of the policy itself, inducing a change in the agents' behavioral rules (\emph{e.g.}, rules on how workers select their effort, how households decide their  savings or how firms invest).

In this section, we elaborate on the third source of parameter instability. Although this issue was raised in 1976 by the celebrated `Lucas critique' of policy evaluation with macroeconometric models analyzing fiscal/monetary regimes, the problem is relevant to any statistical model describing economic phenomena. Under this broader perspective, the diagnosis of the source of parameter variability may be very different to the one posited by Lucas. In other words, unstable parameters are not necessarily related to the individuals' expectations and how they change as a consequence of the announced policy but, instead  -- as earlier suggested by Mises -- , to other factors which cause humans to change their actions. Based on these alternative explanations, we propound a different theoretical framework on how to obtain more stable empirical models through agent computing. Consequently, these models are not built on neoclassical assumptions, but on cognitively-realistic micro-foundations of individuals' behavior and the socioeconomic structure that embeds them.

\subsection{Why decision rules become endogenous}

First, let us clarify the meanings of different terms that are commonly used across the social and behavioral sciences. An individual's action (or decision) is the result of a choice derived from a decision rule (or strategy).\footnote{Although it can also be the product of a social norm, habit, routine or instinct.} A decision (or behavioral) rule is a function that maps a set of variables  -- identifying incentives and perceptions --  to an action. These rules characterize decision-making processes based on maximization principles (\emph{i.e.}, cost-benefit analysis) or behavioral heuristics derived from various learning mechanisms. A decision rule is defined as `exogenous' if its actions are derived from observed incentives (intrinsic and extrinsic) and a given perception of the world. In contrast, a decision rule is `endogenous' when the perception of the world is a function of incentives. The latter scenario occurs, for instance, when a policy modifies the cost-benefit structure, and also the perception of how the world operates. Under these circumstances, it is said that the decision rule shifts (or changes) when facing an intervention. Finally, an individual's behavior is the process by which the decision rule leads him or her to make a decision (or take an action). 

More formally, the behavior of an individual can be described through decision rules

\begin{equation}
  a(p) =
    \begin{cases}
      a(p , \omega)\\
      a(p, \omega(p))
    \end{cases} 
    ,\label{eq:decision}
\end{equation}
where $a(\cdot)$ is the decision rule, $p$ is a policy vector (or set of incentives) and $\omega$ is the perception of the world. This perception relates to how the individual thinks that the world works. This may be informed by signals from policymakers, by the actions of other individuals or by the individual's self-awareness (introspection or self-reflection that makes a person conscious of his or her motivations, virtues, faults, desires and beliefs).

On the right-hand side of equation \ref{eq:decision}, we have two descriptions of behavioral rules, both being functions of the policy vector and the perception of the world. The difference between them is that, in the bottom rule, the individual's perception is also a function of the policy vector. Thus, in the top rule we say that a new policy only affects the actions of the individual due to straightforward material incentives. In contrast, in the bottom one, we say that the same policy may lead to a different choice than the one expected under a static cost-benefit analysis because the decision rule can shift. These rules become endogenous through three major channels: expectations, preferences and social learning. 

When a person faces a new policy, his or her actions may change as a result of observing a different incentive structure. Ultimately, this is the reason why governments decide to intervene in economic systems. In the simplest case, individuals change their decisions based solely on cost-benefit considerations. For example, individuals reduce their fuel consumption in response to a new tax due to a combination of substitution and income effects. However, with  endogenous decision rules, individuals may arrive at a much lower level of fuel consumption. On one hand, the direct effect of incentives remains because this commodity became pricier while, on the other, there is an indirect effect since they do not perceive the world as it used to be. We will now explain why, in this example, a change in perceptions takes place.

There are three major explanations for why, in the previous example, a policy may change our perception of how the world operates. First, the policy may send an ominous signal and, therefore, individuals decide to hold extra precautionary savings (the expectations channel).\footnote{An alternative is that an increase in fuel taxes induces individuals to forecast a reduction in the public deficit and, thus, inflationary pressures will be ameliorated.} Second, a tax on fuel elicits environmental awareness and, hence, people prefer to be more energy-efficient (the preference channel). Third, once implemented, the tax generates dissatisfaction with the government, which resonates and grows as individuals interact and experience the hike in prices. Consequently, fuel consumption might drop even further when a collective movement emerges to boycott the government by `paralyzing' the economy (the social learning channel).\footnote{In comparison with the preference channel, the social learning channel involves feedback loops and a slower process of behavioral change. Moreover, through the social learning channel, we can relate the presence of conformity to the emergence of new preferences in society (\emph{e.g.}, widespread acceptance of a new practice such as vegetarianism). However, it may also reflect the psychological urge to go with the average, without implying any change in preferences.} Under any of these three scenarios, a new perception of the world triggers a shift in the decision rule. If this is the case, a statistical relationship between price and consumption estimated with observational data  -- where no increase in taxes was experienced --  provides misleading policy advice with regard to the consequences of taxation.

\subsection{The social learning channel}

In the context of social learning, individuals respond as a consequence of the collective behavior induced by a policy. In contrast with building expectations from a policy announcement or updating decision rules because a policy appeals to particular preferences, the social learning channel exploits the social nature of individuals who make decisions in terms of reference points and norms.\footnote{For a discussion on the expectation and preference channels see an earlier draft of this paper in \cite{castaneda_evaluating_2018}
and \cite{bowles_economic_2012,bowles_moral_2017}.} For instance, in view of collective rejection of obesity, some people change their eating habits. Other examples include saving more when others do so, becoming an entrepreneur when the startup scene flourishes, participating in public protests when the crowd reaches a critical mass, and becoming more honest when risking exposition under a low-corruption norm. In all these cases collective behavior generates a norm that shifts individual behavior.

A norm does not appear spontaneously, it takes time for it to emerge or for an individual to learn about it. During this process, individuals' decision rules are constantly shifting, until a decentralized understanding on how the world operates is reached. When we speak about norms and conformity, we do not refer exclusively to the process of reaching consensus and settling in a norm (or norms). In this paper, we use these terms in a broader sense, for example, we can say that a high rate of vegetarianism is a norm, not because a broad consensus was reached at the end of the sampling period, but because, on average, a high proportion of the population has adopted a vegetarian conduct through time (even if the rate exhibits large fluctuations or high turnaround rates of adoption). In order to exemplify the social learning channel, we study the process through which a group of people conform to a norm of corruption. This situation is particularly salient in the context of policy prioritization for economic development.\footnote{In the empirical literature there are several studies that analyze the relationship between the social norms of corruption and economic development \citep{mauro_corruption_1995,bardhan_corruption_1997,dutta_corruption_2016}.}  Here, the implementation of public policies is subject to -- socially reinforced -- malpractices such as the diversion of public funds; hindering policy efficacy and overall development.

The problem of formulating and implementing development policies is intimately related to policy prioritization. Here, we can think of a central authority who allocates resources to a group of functionaries in charge of implementing public policies. Each official $i$ receives $P_i$ resources from the government in order to implement a policy that is intended to improve an indicator $I_i$ that measures the level of development in policy issue $i$. This allocation gives the functionary an opportunity to divert a fraction $D_i \leq P_i$ for personal gain. Hence, $C_i = P_i - D_i$ is the effective use of resources in policy issue $i$ (\emph{i.e.,} the contribution of the official). 

How large should $C_i$ be? The answer depends, among other factors, on the level of corruption prevailing in that country at that point in time. For example, a diversion of public funds that is way above the average level of corruption will be more easily detectable by the government's monitoring efforts or by the media, triggering a scandal and exposing the corrupt official. However, this functionary may have passed undetected if his or her transaction had occurred in a situation with a much higher average level of corruption. Therefore, inferring the level of overall corruption allows the functionary to assess the likelihood of being caught.

\begin{figure}[ht!]
    \centering
    \includegraphics[scale=.4]{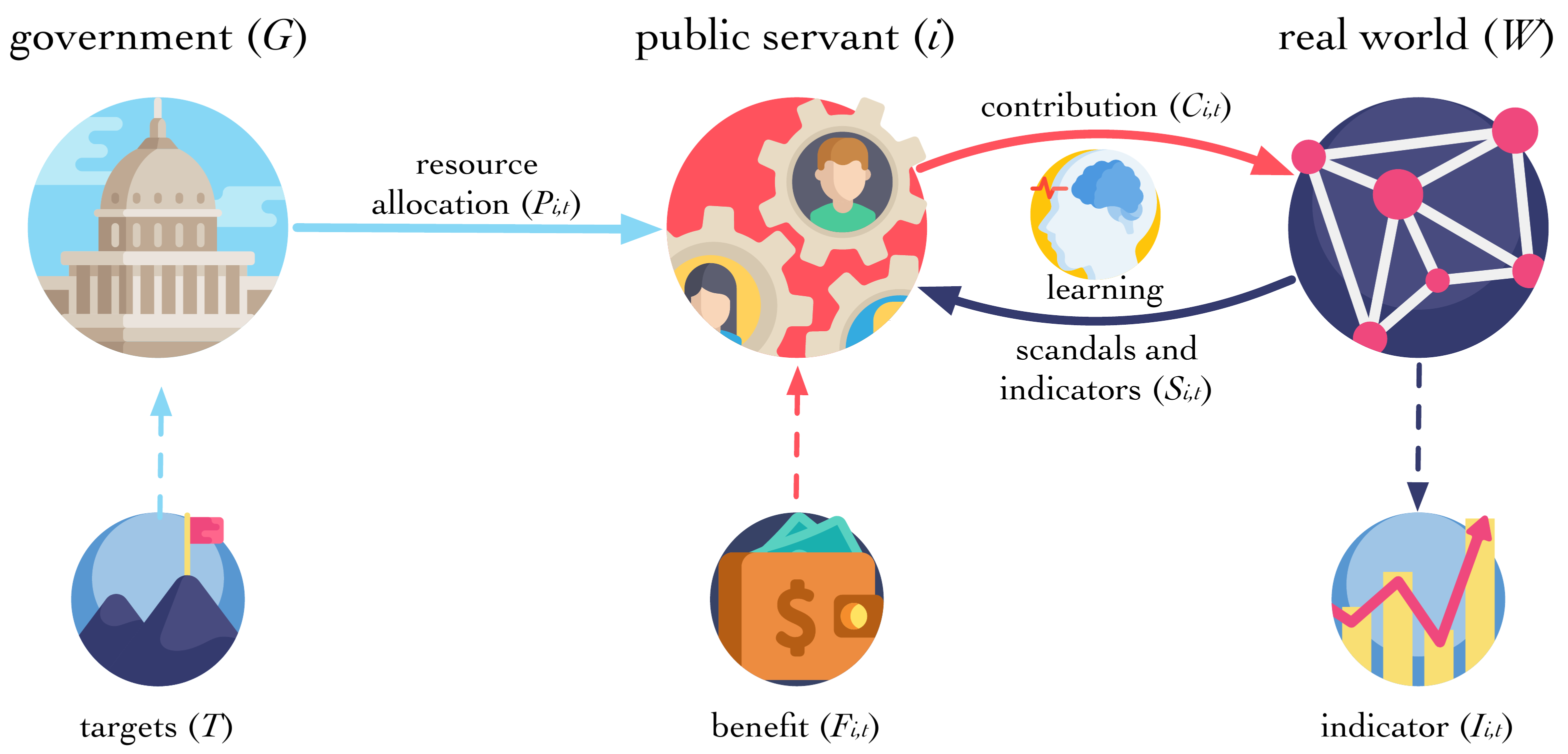}
    \caption{Functionaries' endogenous decision rule through social learning}
    \label{fig:decision}
\end{figure}

Figure \ref{fig:decision} provides a graphical representation of the social learning channel. In period $t$, government $G$ establishes an allocation profile $P_t$ (policy priorities) with the purpose of attaining a specific set of targets $T$ associated to a development strategy. The resources $\emph{P}_{i,t}$ received by public servant $i$ determine his or her contribution $\emph{C}_{i,t}$ which, in turn, affects the real world $W$. Then, the economy's aggregate performance sends tangible signals $S_{i,t}$ to each functionary in terms of the development indicator $\emph{I}_{i,t}$ and media scandals. Next, the official responds to these signals with a new contribution, in order to improve his or her benefits $\emph{F}_{i,t}$. This adjustment continues as functionary $i$ learns the level of corruption that is tolerable by society (or the bearable amount of penalties), giving rise to the norm of corruption. 

In this example, the decision rule for the functionary $i$ is defined as follows: $C_{i,t}=C(P_{i,t}  , \omega(I_{i,t}(P_t) , \rho(P_t)))$, where $\omega$ denotes how he or she perceives the world in terms of his or her development indicator $I_{i,t}$ and the probability $\rho$ of getting caught  which, in turn, depends on the functionary's relative corruption induced by the allocation profile. Since average corruption and indicator levels are a function of the vector of budget allocations, functionaries' decision rules shift when a new policy vector is established, and keep adjusting through time; even if the policy remains fixed.

\subsection{From the economics of control to endogenous governments}

In problems related to economic development, where an active role of the government is generally expected, there is an additional layer of complexity: \emph{endogenous government behavior}. That is, in order to account for the policymaking process, it is necessary to formulate models where the government is an active part of the system. In particular, when the social learning channel matters, specifying a passive or an adaptive government may lead to significant differences in the prescription of a policy. 

In this view, the optimal control program is misplaced. This is so because, in general, real-life policymaking processes do not take place in a Stackelberg setting.\footnote{For an historical review on this program see \cite{colander_complexity_2014}.} In such formulation, policymaking obeys a particular political economy structure where the government passively defines optimal policies, even if the agents' decision rules are considered when formulating policies. In reality, governments are dynamic and adaptive, they react to the decentralized responses that consumers and firms give to such interventions. 

Following an `adaptive' view, policymaking entails a distinction between design and implementation because the government is not a monolithic entity (\emph{e.g.}, there are federal, state, and local levels; legislative and executive branches; technocrats and bureaucrats; boards of advisers and public officials; etc.). These elements give the problem a more dynamic flavor, one where policies are adapted through a sequence of feedbacks originated from the agents' reaction to certain policies (\emph{e.g.}, how the budget is allocated across sectors). The central authority responds to these reactions by updating its policy, most of the time with the aim of attaining certain targets (\emph{e.g.}, punishing corrupt officials/agencies by reallocating the budget). Moreover, in a world with uncertainty, in which individuals cannot be fully rational, learning prevails. This has to be considered in any model that aspires to provide empirically relevant policy advice.

\begin{figure}[ht!]
    \centering
    \includegraphics[scale=.4]{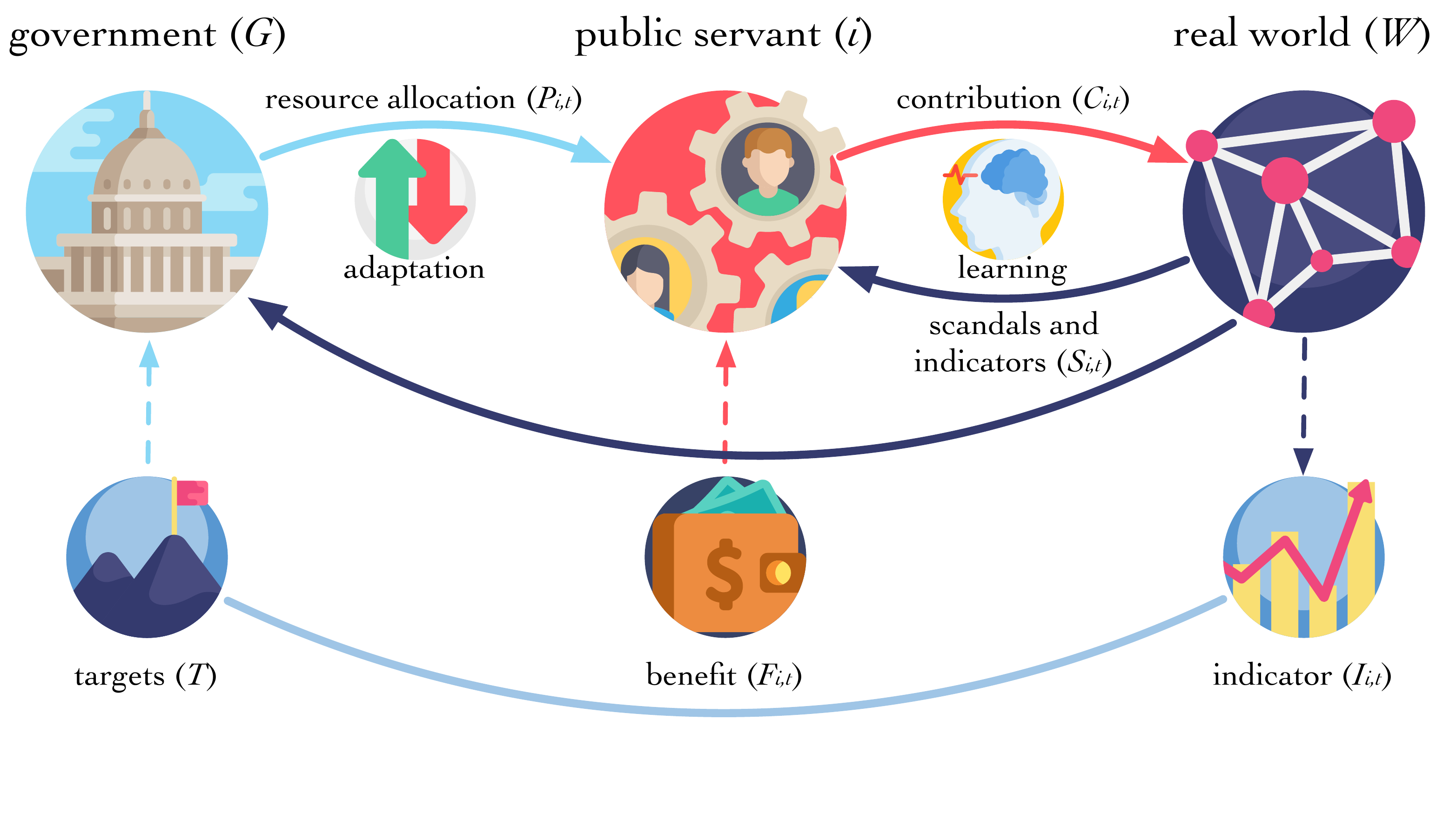}
    \caption{Endogenous government through co-evolutionary learning}
    \label{fig:government}
\end{figure}

Let us elaborate on the dynamic and adaptive quality of the government in the context of the conceptual model described in the previous section. Here, the central authority defines a policy vector $P_t$ without knowing in advance how exactly the agents (in this case, the functionaries) will respond. Instead, it adapts the policy vector according to the performance of the development indicators, and in relation to the levels of corruption inferred through monitoring efforts and media scandals. In the latter form of adjustment, the government reduces budgetary resources for a specific policy issue when its implementation is shadowed by scandals. Therefore, the functionaries' learning process, together with the government's adaptive behavior, gives place to co-evolutionary dynamics, which we illustrate in Figure \ref{fig:government}. In contrast with Figure \ref{fig:decision}, we have included an additional arrow from the economy $W$ to the government $G$, indicating that the central authority adjusts its policies in response to observed variables (the indicators).

In this characterization, the signals that are visible to the government are the same ones observed by public functionaries: scandals coming from corruption and the evolution of development indicators. However, the central authority has at its disposal more information than the functionaries; for example, it knows how close the indicators are from their targets  -- the bottom line connecting $T$ and $I_{i,t}$. This information is used with the purpose of reallocating resources as a response to the functionaries' performance (in terms of their respective indicators $I_{i,t}$) and the diversions of public funds (\emph{e.g.}, by diminishing the available resources if there are signs of corruption).

\subsection{Micro--macro founded models for policy evaluation}

One of the main challenges of dealing with co-evolutionary dynamics is that aggregation becomes non-trivial. Therefore, micro-foundations that rely on homogeneous and rational agents are unfit. A more adequate formulation should consider a richer micro--macro specification. This is so because behaviors are embedded in a socioeconomic context. In terms of model specification, such embeddedness is captured through parameters that represent social norms (\emph{e.g.}, parameters of inter-temporal substitution), collective beliefs (\emph{e.g.}, a law-abiding culture) and economic structure (\emph{e.g.}, productive capabilities). Consequently, the values of such parameters can be endogenized within the model by including the social mechanisms that explain their formation.\footnote{In this respect, \cite{Hoover_microfoundations_2009} argues that some `deep parameters' may be positioned in the social and institutional setting where individual decisions are made.}

It is important to clarify why a model that is capable of accounting for the different types of causal chains in socioeconomic relationships ($macro \rightarrow micro$, $micro \rightarrow micro$ and $micro \rightarrow macro$) does not lead to the inclusion of more unstable parameters.\footnote{For some social scientists, and especially those associated to the research program of analytic sociology, a sensible form to understand the causal relationship between two aggregate variables (\emph{i.e}., financial development and growth) is to study the chain of social mechanisms connecting them (or macro--micro--macro links) \citep{hedstrom_social_1998}. Therefore, instead of imposing axiomatic micro-foundations in a model, a more ontologically-sound approach is to propose alternative chains in agent-based models (ABMs). In contrast to the axiomatic approach, the micro--macro foundations of an ABM are subjected to empirical validation. Should the latter be unfeasible, then these social mechanisms can be internally validated via Monte Carlo simulation \citep{grabner_methodology_2015}.} On the contrary, solid micro--macro foundations allow substituting potentially shifting micro-parameters for stable functions of variables that emerge from collective behavior. 

Models with an explicit micro--macro link are more reliable for policy evaluation because they are less dependent on the stability of estimated parameters. For this to be possible there has to be certain invariance at the level of the social mechanisms involved, which is a more reasonable assumption. Therefore, in these models, changes in observed variables   -- including those related to the formulation of policies -- lead to the endogenous construction of new objects through stable social mechanisms. These objects substitute some of the fixed parameter values in models which are exclusively micro-founded.

In this paper, we provide an example on how agent computing, and the specification of macro-micro links, can be used for parameter endogenization. For instance, in the model presented, the probability of `catching' corrupt officials is defined in terms of a function of endogenous variables. Empirically estimating this probability is problematic because 1) monitoring efforts are a function of the norm of corruption (\emph{e.g.}, a norm where corruption is tolerated implies a government that is lenient to these activities) and 2) because, even with comprehensive data on corruption cases, those who succeed at diverting funds are not observable. The model employed in this paper overcomes these limitations by specifying this probability as a social object that changes endogenously. Then, as long as the assumed mechanism of media scandals is stable, it can be said that the model is reliable for evaluating different policy vectors.

\section{A model of policy priorities}\label{sec:model}

In order to study policy evaluation under social learning with an endogenous government, we use the model by \cite{castaneda_how_2018} (hereafter referred to as CCG), addressing the problem of prioritizing development policies. The CCG model simulates the discovery process of policy prioritization through a behavioral game with two types of agents: a central authority (government) and public officials (functionaries). First, the government allocates resources to different functionaries with the aim of improving the indicators associated to their respective policy issues. Second, the public officials have incentives to divert the assigned funds for personal gain. This game takes place on a network that captures spillover effects between policy issues. These spillovers encourage free-riding and reinforce a misalignment between the government's and the functionaries' incentives. In this section, we provide a brief description of the model.

\subsection{Development indicators}

The economy has $N$ policy issues, each one with an indicator measuring its level of development. The government invests $P_i \in [0,1]$ resources in policy issue $i$ with the purpose of improving the corresponding indicator. This means that a target $T_i$ for policy issue $i$ is always reachable for enough periods and $P_{i} > 0$. We can model the evolution of $I_i$ as 

\begin{equation}
        I_{i,t} = I_{i,t-1} + \gamma(T_i-I_{i,t-1})\left( C_{i,t} + \sum_j C_{j,t}\mathbb{A}_{ji} \right)
    ,\label{eq:propagation}
\end{equation}
where $C_{i,t} \in [0,P_i]$ is the amount of effective resources in policy $i$, $\mathbb{A}$ is the adjacency matrix of the spillover network (a weighted directed graph), and $\gamma$ is a structural parameter reflecting the effectiveness of policy implementation in a given country.

\subsection{Social learning}

Public officials obtain utility from two sources: the level of development of their respective policy issues and diverting public funds. The former gives political status to the functionary, while the latter represents a material gain. In addition, the official may lose utility if he or she gets caught diverting funds. We describe this through the benefit function

\begin{equation}
    F_{i,t} = (I_{i,t} + D_{i,t})(1 - \theta_{i,t}f_{R,t}),\label{eq:benefit}
\end{equation}
where $\theta_{i,t} \in \{0,1\}$ is the outcome of the government's monitoring of corruption and $f_{R,t}$ maps the quality of the \emph{rule of law} into $[0,1]$, denoting how \emph{tolerant} it the government towards corruption (\emph{i.e.}, its capacity to punish misbehavior when detected).

In order to determine his or her contribution, the public functionary follows a directed-learning process dictated by

\begin{equation}
    C_{i,t} = \min \left\{ P_{i,t},
     \max\left(0, C_{i,t-1} +  d_{i,t} |\Delta F_{i,t}| \frac{C_{i,t-1} + C_{i,t-2}}{2} \right) \right\} \label{eq:contribution},
\end{equation}
where $\Delta F_{i,t}$ is the most recent change in benefits and $d_{i,t}$ is the sign function

\begin{equation}
    d_{i,t} = \text{sgn} (\Delta F_{i,t} \cdot \Delta C_{i,t}),\label{eq:sign}
\end{equation}
such that

\begin{equation}
    \begin{split}
        \Delta F_{i,t} = F_{i,t-1} - F_{i,t-2}\\
        \Delta C_{i,t} = C_{i,t-1} - C_{i,t-2}.
    \end{split}
\end{equation}

That is, functionaries' contributions increase when higher (or lower) past benefits coincide with higher (or lower) past contributions.

\subsection{Adaptive government}

We assume that the diversion of public funds is not observable to the government, unless a societal signal makes it stand out (\emph{e.g.}, a successful investigation by the political opposition, leaked documents, corruption scandals in the media, etc.). Since signals can be random, we assume that $\theta_{i,t}$ is a Bernoulli random variable with probability

\begin{equation}
         f_{C,t}\frac{(P_{i,t}-C_{i,t})}{\sum_{j=1}^N (P_{j,t}-C_{j,t})},
\end{equation}
where $f_{C,t}$ is a function mapping the quality of the \emph{control of corruption} into $[0,1]$. Since the probability depends on endogenous variables, $\theta_{i,t}$ follows a time-varying process. Note that the probability of being caught is proportional to the size of the diversion, but relative to the overall amount of diverted resources. This means that the learning process is a function of a systemic property: \emph{the norm of corruption}.

With respect to the maps $f_{R,t}$ and $f_{C,t}$, we can say that both are functions of endogenous variables. These are development indicators in the \emph{rule of law} and the \emph{control of corruption} respectively. The two mechanisms describe different constraints that governments face when fighting corruption. To be more specific, they take the form

\begin{equation}
    f_{X,t} = \frac{I_{X,t}}{e^{1-I_{X,t}}},
\end{equation}
where $X=R$ for the \emph{rule of law} or $X=C$ for \emph{control of corruption}.

Besides monitoring and punishing corruption, the government has to decide how to allocate its limited resources across all policy issues. By establishing a target $T_i$ for each indicator $I_i < T_i$, the central authority aims at closing their gaps. Formally, this multidimensional problem is 

\begin{equation}
    \min \left( \sum^N_{i=1} \left| I_{i,t}-T_{i} \right| \right)  .\label{eq:MSE}
\end{equation}

The allocations $P_{1,t}, \dots, P_{N,t}$ are the control variables of the government. We call a specific configuration $\{ P_{i,t} \}_{i=1}^N$ of these variables an \emph{allocation profile}. The amount of resources that the government can invest per period in a profile is restricted by

\begin{equation}
    \sum_i^N P_{i,t} = B \; \forall \; t.\label{eq:budget}
\end{equation}
where $B$ denotes \emph{non-committed} resources of the central authority. It is important to clarify that the resources involved in this problem are those destined to transformative policies; not public expenditure committed to previously established purposes (\emph{e.g.}, highway maintenance, agricultural subventions, payment of public debt, etc.).

Each time step, the central authority adapts its allocation profile by prioritizing laggard policy issues (in terms of the gap $T_i-I_i$) and exercising budgetary punishments to those functionaries who were found diverting public funds. Then, it allocates resources to policy issue $i$ with propensity 

\begin{equation}
    q_{i,t} = (T_i-I_{i,t})(K_i+1)(1-\theta_{i,t}f_{R,t}),\label{eq:propensity}
\end{equation}
where $K_i$ is the number of outgoing connections of node $i$, also known as its \emph{out-degree}. Here, the out-degree captures the \emph{centrality} or importance of a policy issue for the economy. Hence, the government does not know the structure of the network, but has a proxy of the relevance of a policy issue.

Finally, normalizing for all propensities and the budget constraint, we obtain the allocation

\begin{equation}
    P_{i,t} = \frac{q_{i,t}}{B\sum_j^N q_{j,t}}.
\label{eq:allocation}
\end{equation}

Note that, for each simulation, we obtain four endogenous vectors: public officials' diversions $D$, their benefits $F$, the government allocations $P$, and societal development indicators $I$. A simulation halts when a convergence criterion is met for all indicators.\footnote{In this model, a period represents the realization of some events. For example, achieving a target in $\ell$ periods means that the government had to experience $\ell$ events of budget reallocation. A larger $\ell$ implies that reaching the target was more difficult. Therefore, $\ell$ should not be interpreted in terms of time units.}  The only free parameter in the model is $\gamma$, which is used for a cross-national calibration. All functions describing the model's social mechanisms adjust their values as the simulations run.\footnote{See \cite{castaneda_how_2018} for further details on the motivations, the implementation and the calibration of the CCG model.}

\section{Policy prescription and evaluation}\label{sec:method}

In order to prescribe policy priorities, our methodology considers a two-tier package. The first tier is an estimated allocation profile $\hat{P}_0, \dots, \hat{P}_N$ that reflects the expected prioritization of policies when a government aims to reach a specific set of targets $T_0, \dots, T_N$. The second tier relates to the flexibility that a central authority has when following a recommended allocation profile. In the CCG model, such flexibility translates into budgetary readjustments triggered by corruption scandals (see equation \ref{eq:propensity}). Thus, in a way, strictness implies scarifying adaptability to attend societal pressures.

\subsection{Epistocratic \emph{versus} technocratic advice}

In order to evaluate policy prescriptions, it is useful to frame them in the context of epistocratic \emph{versus} technocratic  advice. A technocrat does not have a systemic understanding of the economy and the policymaking process. Hence, technocratic (uninformed about the systemic nature) policy prescriptions can be described as disarticulated conjectures subject to a budget constraint. In contrast, we say that an epistocrat provides an informed recommendation because he or she \emph{discovers} the allocation profile that would be developed when the government tries to reach a set of targets.\footnote{In the context of this paper, a technocrat is an expert (economist or otherwise) in a specialized field who knows how the economy works in specific areas (\emph {e.g.,} the banking sector, public health, telecommunications and infrastructure), but who is unaware of the linkages between the different dimensions of economic development and the prevailing political economy. In contrast, an epistocrat is an analyst who is knowledgeable of the policymaking process and how it relates to the overall functioning of the economy, in a systemic sense.} Allocation profile discovery is feasible when the epistocrat employs an empirically validated model of the policymaking process, which in our case is an agent-computing model of a political economy game on a network. Strictly adhering to a policy recommendation means that a government has to be disciplined and sacrifice the adaptive quality through which it updates the allocation profile (see equation \ref{eq:propensity}). At the same time, public officials learn only through the evolution of their indicators and the direct penalty to corruption (not through budgetary punishments; see equations \ref{eq:contribution}, \ref{eq:benefit} and \ref{eq:propagation}).

An epistocrat can formulate two types of advice: a \emph{strict-informed policy} or a \emph{lax-informed policy}. The former can be formalized as

\begin{equation}
    \{ P_{i,t}^{I} \}_{i=1}^N =  \{ \hat{P}_i^{D} \}_{i=1}^N  \qquad \text{for every} \quad  t
    ,\label{eq:strict-informed}
\end{equation}
where $P_{i,t}^I$ is the recommended budgetary allocation for policy issue $i$ in period $t$, and $\hat{P}_i^D$ denotes the allocation to policy issue $i$ estimated through the ABM (\emph{i.e.} the discovered allocation).\footnote{We assume that the initial allocation profile in the discovery phase is the arbitrary profile $\{ A_i \}_{i=1}^N$.} 

On the other hand, a lax-informed prescription is given by

\begin{equation}
  \{ P_{i,t}^{I} \}_{i=1}^N =
    \begin{cases}
       \{ \hat{P}_i^{D} \}_{i=1}^N \qquad \text{if} \qquad t=0 \\
       \{ P_{i,t} \}_{i=1}^N  \qquad \text{otherwise} 
    \end{cases}
    ,\label{eq:lax-informed}
\end{equation}
where $P_{i,t}$ is the allocation for policy issue $i$ determined endogenously during the policymaking process. The lax-policy advice indicates that, after using the analyst's initial recommendation, governments should determine their priorities by acting normally (\emph{i.e.}, with the usual adaptive mechanisms that include budgetary punishments).

When adopting a strict-informed advice, governments implicitly face significant political and logistic constraints. This is so because being disciplined implies `giving up' budgetary adjustments as tool to penalize corruption. On one hand, political constraints involve criticisms from the civil society, the media and the opposition for not cutting resources that are susceptible to diversion. On the other, logistics' constraints refer to the costs of reorganizing agencies swiftly in case public officials are removed, so that resources keep flowing. Generally speaking, these constraints impose severe limitations to strictly following $\hat{P}_i^D$; so disciplined implementations might be unfeasible in a practical sense.

\subsection{Benchmark allocation profile}

Our benchmark to evaluate policy prescriptions is the advice of a technocrat (uninformed analyst); in particular, a \emph{lax-uninformed} prescription. Uninformed recommendations can be derived from a vast space of potential policy priorities that might make sense at the policy-issue level, but not necessarily at the systemic one. Therefore, we explore the entire set of potential priorities through random prescriptions.\footnote{An uninformed advice can be interpreted as a collection of partial policy suggestions, derived from disconnected theories and models (\emph{e.g.}, macroeconomic priorities from a DSGE model, health priorities from growth diagnostics, business sophistication priorities through regression analysis, education priorities by benchmark comparison of indicators, and so on and so forth), which are then put together in a discretionary allocation profile.} Formally, a strict-uninformed prescription is given by the following expression:

\begin{equation}
    \{ P_{i,t}^{U} \}_{i=1}^N =  \{ A_i \}_{i=1}^N  \qquad \forall \quad  \emph{t} \quad \text{until convergence} 
    ,\label{eq:strict-uninformed}
\end{equation}
where $A_i$ is an arbitrary allocation. 

Then, for lax-uninformed prescriptions, the government can make adjustments to the allocation profile in every period depending on the signals that the economy generates. These prescriptions are given by

\begin{equation}
  \{P_{i,t}^{U} \}_{i=1}^N =
    \begin{cases}
       \{ A_i \}_{i=1}^N \qquad \text{for} \qquad t=0\\
       \{ P_{i,t} \}_{i=1}^N  \qquad \text{otherwise} 
    \end{cases}
    ,\label{eq:lax-uninformed}
\end{equation}

After inserting these expressions in our algorithm, we use Monte Carlo simulation to compute expected allocation profiles and corruption estimates.

\section{Empirical analysis}\label{sec:empirics}

Our empirical analysis consists of evaluating the outcomes of different allocation prescriptions. This is done by estimating the amount of diverted resources which, for a single simulation, is given by

\begin{equation}
    L = \frac{1}{B}\sum_i \sum_t (P_{i,t}-C_{i,t}),\label{eq:totCorr}
\end{equation} 

Let equation \ref{eq:totCorr} denote the level of corruption in the benchmark case (lax-uninformed), and $L'$ the level of corruption in an alternative case (\emph{i.e.} strict-uninformed, lax-informed and strict-informed). From an aggregate point of view, $L-L'$ denotes \emph{gains (or losses) in efficiency} through a reduction (increase) in  corruption when adopting an alternative prescription to the benchmark.

\subsection{Data}

The data consist of annual observations of 79 development indicators for 117 countries, covering the 2006--2016 period. Three secondary sources are used to build this database: the Global Competitiveness Report produced by the World Economic Forum, the World Development Indicators and the Worldwide Governance Indicators; the latter two assembled by the World Bank. We normalize these indicators so that the worst possible outcome takes a value of zero, while the best one has a value of one across countries and years.\footnote{For a detailed description of this database and all variables used to validate the model empirically, see \cite{castaneda_how_2018}}. 

A spillover network is built for each country using the time series of its development indicators. The method of choice to construct a network is a two-step strategy. First, we apply the method of triangulated maximally filtered graphs (TMFG) \citep{massara_network_2017} to estimate which pairs of indicators have significant relationships. Then, we determine the edges' directions through the likelihood-ratios method developed by \cite{hyvarinen_pairwise_2013}.  

The idea behind our policy evaluation exercises is to simulate the development indicators of each country in the sample period. For the lax-uninformed case, this has already been done by \cite{castaneda_how_2018} through a cross-national estimation that fits the model's endogenous variable of corruption to an empirical indicator of the \emph{diversion of public funds}. Therefore, we can say that the CCG model provides a plausible mechanism to explain the levels of corruption observed in the dataset. By running simulations for lax-informed, strict-uninformed and strict-informed prescriptions, we are effectively performing counterfactual analyses to estimate what would have been the gains or losses in efficiency from these types of policy advice.

As inputs for our simulations, we use the levels of the development indicators in 2006 -- or the most distant year available -- as initial conditions $\{I_{i,0}\}_{i=1}^N$ for each country. For its targets $\{T_{i}\}_{i=1}^N$, we use the levels obtained in 2016 -- or the most recent annual data available for the indicators. The adjacency matrix $\mathbb{A}$ of the spillover network is estimated from the country's indicators, and its budget constraint $B$ is obtained from public expenditure as a fraction of GDP.

\subsection{Results 1: distributions of corruption across countries}

\begin{figure}[ht!]
    \centering
    \includegraphics[scale=.5]{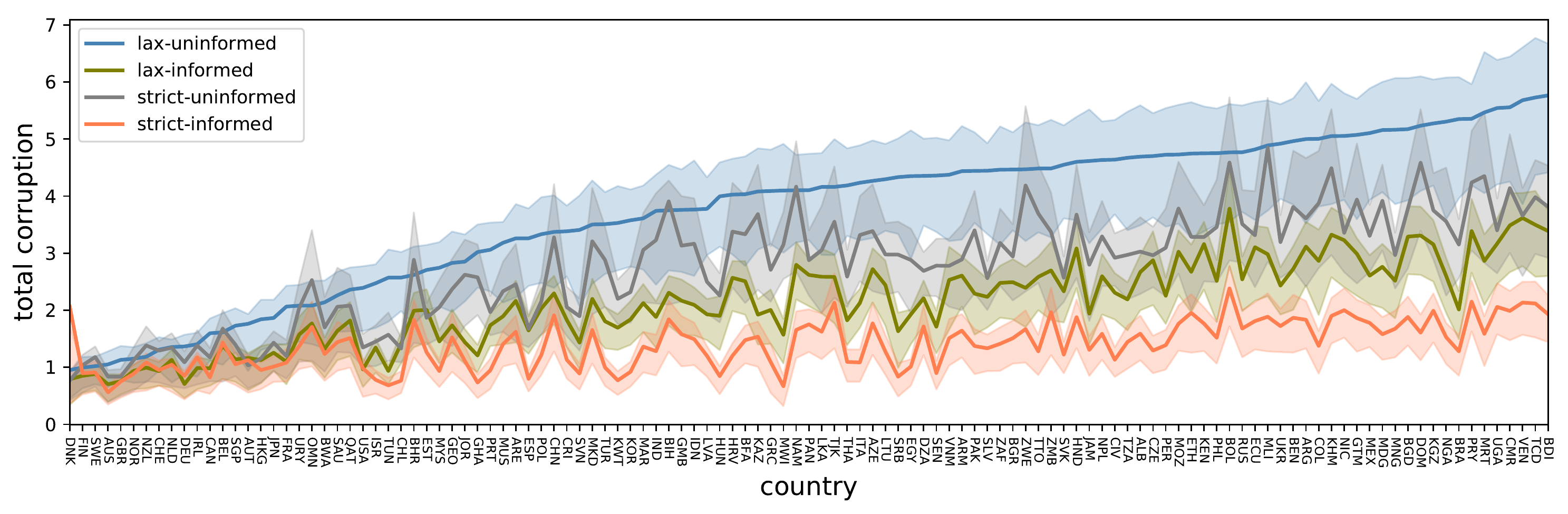}
    \caption{Model outputs by country (sorted by level of corruption under the benchmark). The shaded areas denote the region from the $25^\text{th}$ to the $50^\text{th}$ percentiles of each distribution. The solid lines indicate the sample mean. We perform 1000 simulations per country and per type of policy}
    \centering \label{fig:country_level}
\end{figure}

We investigate if differences in corruption between policy prescriptions vary across countries. Figure \ref{fig:country_level} shows the corruption distributions between the $25^\text{th}$ and the $50^\text{th}$ percentiles for each country and each type of advice. Clearly, they vary considerably across nations, highlighting the importance of context-specificity. Here, countries have been sorted by the level of corruption under the benchmark (blue solid line). In average terms, the three counterfactual policies perform better than the benchmark. Likewise, the plot indicates that there are overlaps between distributions in most countries, and this is particularly large among low-corruption nations (\emph{i.e.}, the most developed ones). Accordingly (and supported by $t$-tests), we can assert that the efficiency gains of alternative policy prescriptions are statistically significant for many countries, especially those with medium and high levels of corruption.

The simulation results also indicate that strict-informed policies (orange line) produce the lowest average levels of corruption, and informed policies perform better than uninformed ones. Moreover, it is better to sacrifice discipline under an informed policy (green line) than to adopt an uninformed one while being completely strict (gray line). From this \emph{ex ante} policy evaluation, we conclude that the epistocratic advice yields better outcomes than the technocratic one. This result indicates that the suggested policy prioritization leads to a series of societal signals that incentivize the public servants to adjust their contributions towards the establishment of lower corruption norms.

There is one more observation particularly salient in Figure \ref{fig:country_level}. Corruption levels are generally lower under strict policies. This is quite a paradoxical result since strict packages dictate that governments should not readjust their budgetary allocations in the presence of corruption. That is, by limiting the functionaries' costs of infringement, a country becomes less corrupt with the passing of time. An explanation for this intriguing result is that, by reducing the sources of punishment, the frequency of media scandals increases. This, in turn, induces functionaries to conform, unconsciously, to a lower norm of corruption.\footnote{In our model, functionaries are not cognitively aware of the existence of a corruption norm; however, their directed-learning heuristic creates an indirect incentive to `hide' since this action diminishes the possibility of a reduction in benefits.}

\subsection{Results 2: where to find gains in efficiency?}

It is useful to disaggregate the origins of gains in efficiency in terms of the topics covered by each policy. For clarity of exposition, instead of considering our 79 socioeconomic indicators, we analyze the composition of gains in efficiency in terms of 13 commonly used development pillars. In addition, it is also useful to perform this analysis across different sub-samples of countries. This is so because the distribution of gains in efficiency across development pillars may vary significantly between, for example, low-income and high-income countries. Evidence of these differences would indicate the importance of country-specific context in the prescription of public policies.

In order to create sub-samples of countries, we identify four clusters by applying Ward’s method with the L2 (Euclidean) norm as the distance metric across the 79 indicators. These clusters roughly correspond to the four income groups of the World Bank, with the difference that we account for more than one dimension of development. 

\begin{figure}[ht!]
    \centering
    \includegraphics[scale=.5]{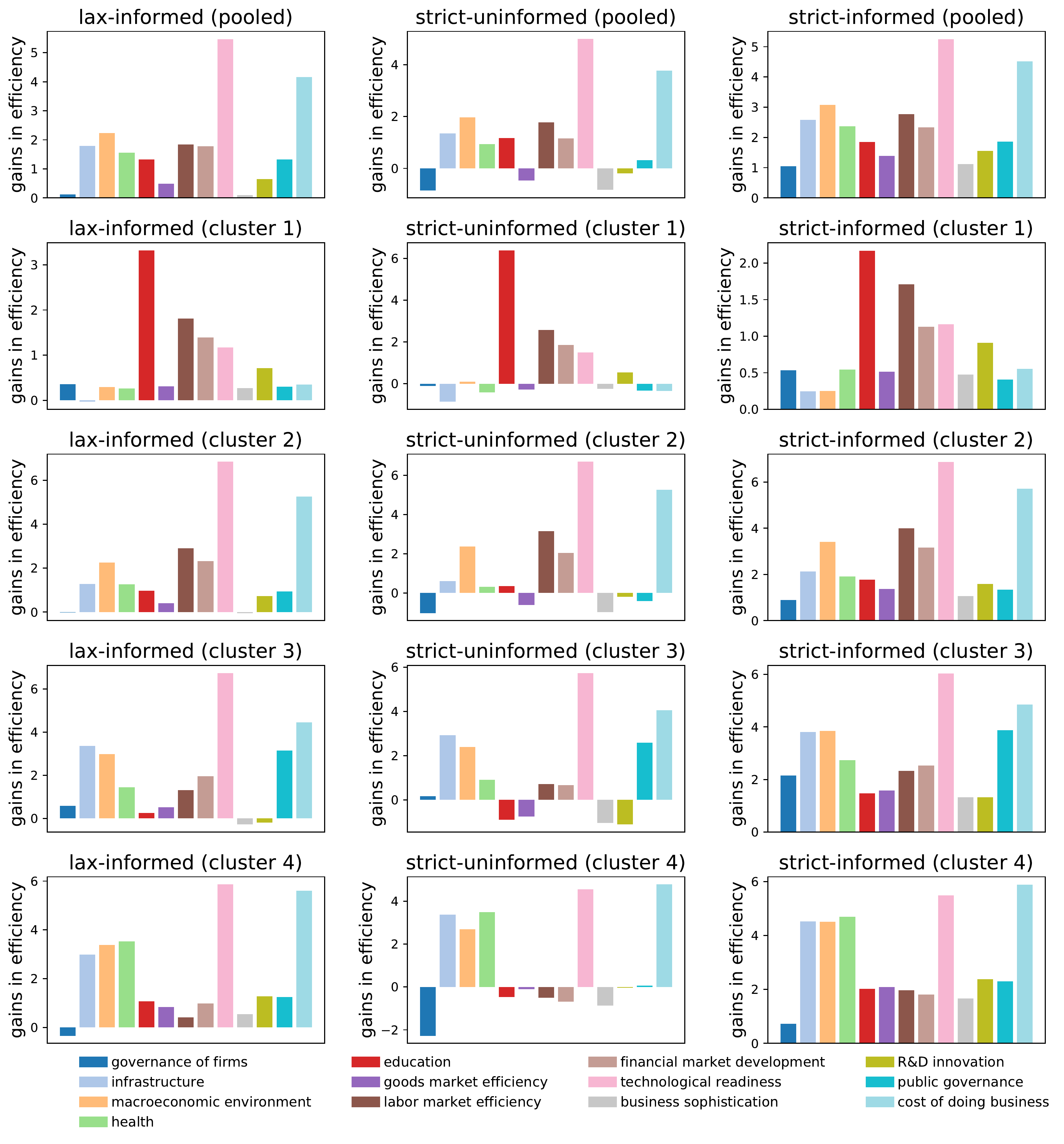}
    \caption{Gains in efficiency by development pillar. Each bar corresponds to the average gains in efficiency with respect to lax-uninformed policies (benchmark). The first row of panels corresponds to the average values of the pooled data.}
    \centering \label{fig:barsBoth}
\end{figure}

According to the different panels in Figure \ref{fig:barsBoth} the distribution of efficiency gains across pillars is very heterogeneous, especially for strict-uninformed advice. As shown in the middle column, prescriptions without information on the discovery of policy priorities produce losses in efficiency across several pillars (\emph{i.e.,} corruption is larger than in the benchmark). When comparing rows, on the other hand, it is clear that the distribution of efficiency gains varies between groups. For instance: gains in \emph{education} are the largest for the average country in cluster 1, irrespective of the adopted prescription, while this is not the case for the other clusters. Likewise, gains in \emph{infrastructure} and in \emph{macroeconomic environment} are negatively related to the level of development,\footnote{The indicators included in the \emph{macroeconomic environment} pillar are: \emph{inflation}, \emph{government debt as a percentage of GDP}, \emph{foreign direct investment} and \emph{imports as a percentage of GDP}. Therefore, corruption in this pillar comes mainly form government expenditures related to the federal bureaucracy, and with policies fostering an open economy.} while \emph{labor market efficiency} exhibits a positive relationship. Then, gains in efficiency from \emph{technological readiness} and \emph{cost of doing businesses} are extremely high for clusters 2-4, but not for cluster 1.

\subsection{Results 3: some country cases}

Instead of using Pareto-efficiency as our driving criteria to discriminate policy prescriptions, we discover admissible profiles by simulating the policy-making process and generating the distribution of corruption. From these, the analysts can select those that tend to exhibit gains in efficiency with respect to some benchmark. In other words, policies are not designed deductively to meet certain theoretical conditions that appeal to axiomatic preferences and perfect rationality. Instead, they are found by studying the dynamics of an `artificial economy' that resembles the real one.\footnote{One can evaluate these policies in terms of a richer and more flexible set of criteria that are more relevant to the particular problem at stake (\emph{e.g.}, corruption, inequality, social inclusion, ecological sustainability).}

\begin{figure}[ht!]
    \centering
    \includegraphics[scale=.53]{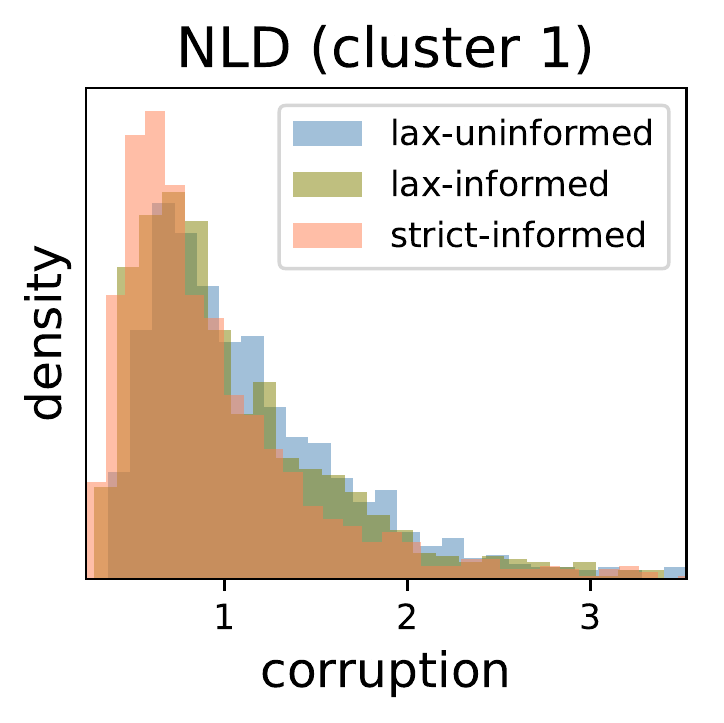}
    \includegraphics[scale=.53]{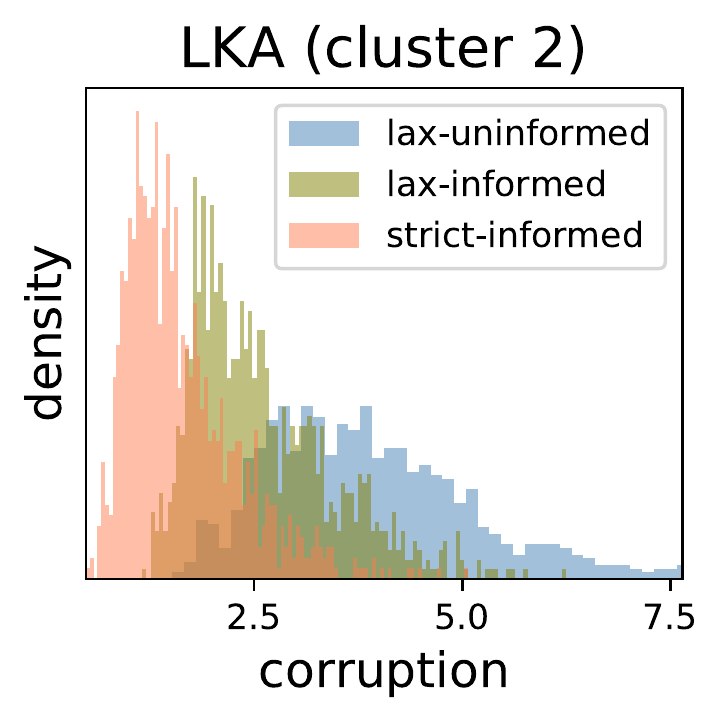}
    \includegraphics[scale=.53]{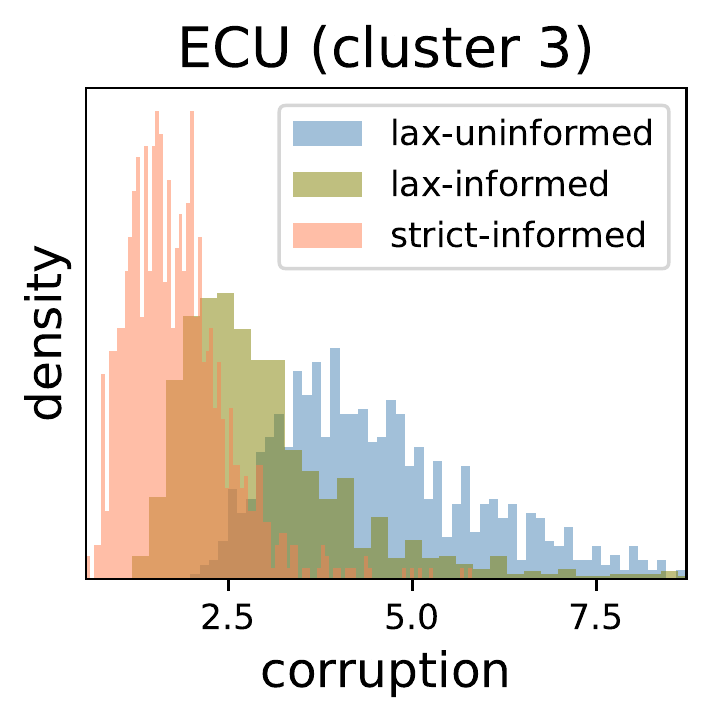}
    \includegraphics[scale=.53]{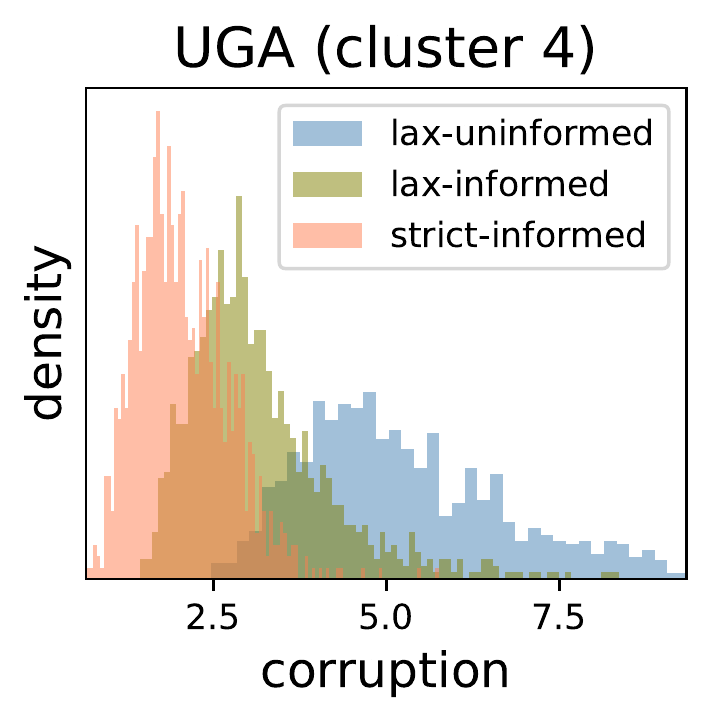}

    \caption{Distributions of corruption under informed policies compared to the benchmark (lax-uninformed). Examples with significant overlaps.}
    \centering \label{fig:mechanismDesign}
\end{figure}

For illustration purposes, we analyze the distribution of corruption outcomes for one country of each cluster. In particular, through Figure \ref{fig:mechanismDesign}, we study how often epistocratic advice yields efficiency gains with respect to the benchmark in our simulations. As mentioned previously, the average performance of strict-informed policies is better than the benchmark's. However, for the country cases selected here (Netherlands, Sri Lanka, Ecuador and Uganda), there are clear overlaps between the distributions from informed prescriptions (orange and green) and lax-uninformed ones (blue). This result implies that even if informed policies are better on average, some countries have a significant likelihood of obtaining outcomes similar to the expected ones under uninformed prescriptions.

From these examples it follows that, as long as these distributions present overlaps, we cannot talk about Pareto-efficient policies.\footnote{A lower level of corruption not only makes society better off, but also can improve the bureaucrats' utility. That is, public officials compensate their lower personal gains from corruption with higher political status due to improvements in development indicators.} That is, the same policy prescription may lead to various levels of corruption across the different simulations; all equally parameterized. Hence, a strict-informed prescription can be Pareto superior with respect to the benchmark when evaluated at mean values, but it can also be inferior for particular simulation runs. We conjecture that this ambivalence stems from the multiplicity of corruption norms that can emerge under the same policy package. 

This outcome highlights the importance of path-dependence through the learning process and the discovery of policy priorities. That is, in a realization of the world, a bureaucrat may be caught diverting funds (recall that monitoring is a random event), while in a different realization he or she may succeed. Depending on the order and synchronization of these events, the population may be locked into paths of lower or higher corruption norms. In other words, observed outcomes result from bottom-up processes (and random elements) that lie beyond the policymaker's control.

When we move from lax-uninformed to lax-informed prescriptions there is a significant shift of the latter distribution to the left (confirmed by \emph{t}-tests) for the four country cases. Presumably, this implies that initial allocations originating from an exploratory phase `help' the system to establish lower corruption norms. This result, in turn, is a consequence of initial societal signals that discourage functionaries from stealing. Finally, the domain of the distribution shifts even more when we move from lax to strict prescriptions, especially for middle and low-income countries.

\section{Discussion on policy advice and evaluation}\label{sec:discussion}

The results presented in this paper provide new insights into the limitations of technocratic policy advice. It is well known that policy prescriptions from international organizations such as the World Bank and the IMF are often criticized from many angles. In fact, cases of stagnated economies such as Latin-America are commonly used as an example of the ineffectiveness of such recommendations. In response, supporters of these prescriptions argue that such failures have nothing to do with the recommendations, but rather with poorly disciplined adoptions. Our study sheds light on an important issue that has been previously ignored: that poor policy performance may have less to do with lack of discipline and more with the non-systemic quality of the prescription (\emph{i.e.}, its uninformed nature). Therefore, a systemic approach to understand the formulation of policy priorities could help conciliating both sides of this debate.

Another interesting angle in this debate is that, as our results show, the difference in outcomes between informed and uninformed policies seem negligible among developed nations while significant in developing ones. If one considers that much of the policy advice coming from multilateral organizations is based on the experience of industrialized economies, it should not be surprising to observe a disregard for the context in which the policy is supposed to be applied. Thus, our study 1) supports the long standing criticism to generic policy recipes and 2) exhibits the problem of building general theories from a narrow empirical experience.

Another topic in which our approach can shed new light is \emph{policy coherence}, a concept that has recently gained popularity among multilateral organizations and academics who acknowledge the multidimensionality of development. This concept usually involves identifying trade-offs and complementarities in a large set of policy issues. These (trade-offs and complementarities), in turn, cannot be fully understood without a systemic approach. Thus, current methods that assess policy coherence by comparing development indicators to priorities officially stated in government documents can be highly misleading. In contrast, our study suggests that a more correct approach should compare inferred priorities during the sampling period (using the empirical targets $T$) with those estimated from a counterfactual exercise in which $T$ corresponds to the officially stated development goals. In this way, it would be possible to provide a quantitative measure of policy coherence built on systemic considerations of the policymaking process.

The impossibility of performing ex ante evaluation directly from development indicators extends to public expenditure data. This is so because, for example, the total resources spent in improving, for example, \emph{public health} do not reflect the diversions and readjustments that took place in the sampling period (not to mention the spillover effects). Furthermore, if the data is relatively aggregate (like most public finance data is), it would not distinguish between resources devoted to maintain the current infrastructure of, for instance, public hospitals from those destined to transform the sector. This problem is especially severe among developing countries, whose records on public finance tend to be produced as a mere formality rather than a commitment to fiscal transparency. For these reasons, policy evaluation that employs this type of information should be careful about ignoring systemic features.

By discovering allocation profiles that generate gains in efficiency, our study warns about the risks of over-promoting certain development agendas. For instance, while \emph{public governance} can be an instrument to mitigate corruption, our results suggest that there are many allocations $P$ where similar gains can be obtained by prioritizing other type of indicators. Thus, over-emphasizing certain topics in a development strategy can turn policymakers blind to important complementarities from other issues.

As a final thought, we would like to emphasize that this framework can help reducing dependence on expert advice. While such advice is extremely helpful for policy design, the reality is that it tends to be a scarce (or even unaffordable) resource in many developing countries, especially at sub-national levels. In numerous countries, elaborating development plans is mandatory at state/provincial levels. Thus, in absence of expert advice, evidence-based policy is rarely used. This has caused benchmark comparisons to become the norm; although, some other discretionary criteria can be common too. Unfortunately, and for the same reasons discussed above, these approaches are ill-suited for policy evaluation. Thus, policy failure may be around the corner despite having well-intended governments.

\bibliography{references3}

\end{document}